\documentclass[prb,twocolumn,aps,showpacs]{revtex4-1}
\usepackage{graphicx}% Include figure files
\usepackage{color,latexsym}
\def\be{\begin{equation}}
\def\ee{\end{equation}}
\def\ds{\displaystyle}

\begin{document}

\title{ Comment on the paper  \\
I. M.  Suslov: Finite Size Scaling from the Self Consistent Theory of Localization}

\author{P. Marko\v{s}}
\affiliation{Department of Physics FEI, Slovak University of
  Technology, 812\,19 Bratislava, Slovakia}

\begin{abstract}
In the recent paper [I.M.Suslov, JETP {\bf 114} (2012) 107]
a new scaling theory of electron localization was proposed. We show that  numerical data
for the quasi-one dimensional  Anderson model do not support predictions of this theory.
\end{abstract}


\maketitle

\section{Introduction}

In the  recent paper \cite{1},   the scaling theory of electron localization is discussed.
It is   argued that the standard  interpretation of numerical data
based on the finite size scaling analysis \cite{sc,so,2}  is not correct.
For the quasi-one dimensional Anderson model, new formulation
of the scaling, based on the analytical self-consistent theory, is presented. The theory      
gives  for the three dimensional (3D) Anderson model  the critical exponent $\nu=1$,
in agreement with original self-consistent theory of Anderson localization  \cite{vw}.
New scaling relations have been proposed for higher dimension $d>4$. 

In this comment we show that 
the theory\cite{1}  is not consistent with  present numerical data
for the 3D and 5D Anderson model. 

We consider Anderson model\cite{a} with diagonal disorder $W$ 
defined on the quasi-one dimensional system of the size
\be
L^{d-1}\times L_z~~~~~L_z\gg L
\ee
($d$ is the dimension of the model) and calculate the smallest Lyapunov exponent $z_1(W,L)$. The later is related to the localization length
$\xi_{\rm 1D}$
\be\label{eq-z1}
z_1 = \frac{2L}{\xi_{\rm 1D}}
\ee
and determines the exponential decrease of the wave function,
$|\Psi|^2\sim\exp [-z_1 L_z/L]$. \cite{2}
For the 3D model, $L_z=2 L/\varepsilon^2$  is sufficient to achieve the relative  
numerical accuracy
$\varepsilon$. \cite{jpcm} The size $L$ varies from $L=8$ to $L=34$ for $d=3$ and
is $L\le 8$ for $d=5$. 

\section{3D system}

Suslov's theory predicts  \cite{1} 
that in the vicinity of the critical point
($\tau = W-W_c\ll 1$)
the localization length follows the scaling behavior 
\be\label{s-1}
\frac{\xi_{\rm 1D}}{L} = y^* + A \tau (L+L_0)
\ee
with a new  additional length scale $L_0$ not considered in the standard scaling analysis.
($y^*$ is the size-independent critical value).
This prediction is in variance with the standard scaling formula
\be\label{s-1a}
z_1 = \frac{2L}{\xi_{\rm 1D}} = z_{1c}+ A\tau L^{1/\nu},
\ee
used in the finite size scaling analysis of numerical data. \cite{sc,so}

\begin{figure}[b!]
\includegraphics[width=6cm]{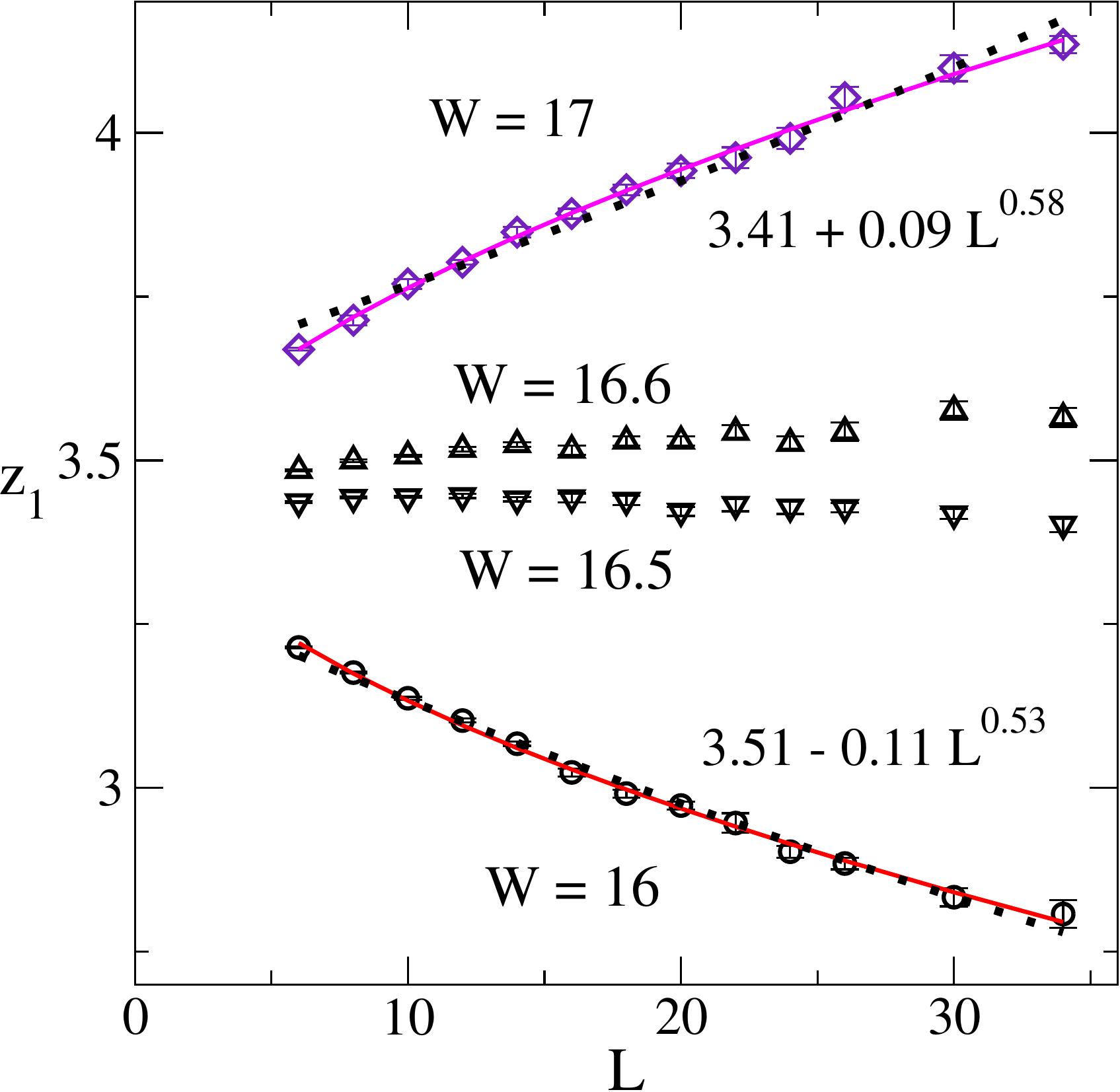}
%\vspace*{0.5cm}
\caption{The 3D Anderson model: 
The parameter $z_1(L)$ for various disorder 
Solid lines are power fits for $W= 16$ and $W=17$.  Contrary to \cite{1},
fits are not linear in $L$.
Note that $z_1$ decreases for $W=16.5$ and increases for $W=16.6$. Therefore, we expect that
$16.5<W_c<16.6$. Scaling analysis gives $W_c\approx 16.55$.
Dotted lines are fits (\ref{fit-s}) with $a_0=0.302$ and $a_1=0.0017$ ($W=16$) and
$a_0=0.267$, $a_1=-0.00108$ ($W=17$). 
}
\label{w16-17}
\end{figure}

To support the result (\ref{s-1}), Suslov used  numerical data for parameter $z_1$ 
published in Ref. \cite{2} and  found  that $L_0\approx 5$ (left Fig. 6 in \cite{1}).
We show in Fig. \ref{w16-17} the same Figure  with additional data for $24\le L \le 34$.
Power fit $z_1(L) = a + bL^\alpha$ calculated for $W=16$ and $W=17$ supports
the validity of the the relation (\ref{s-1a}).

\begin{figure}[t!]
\includegraphics[width=6cm]{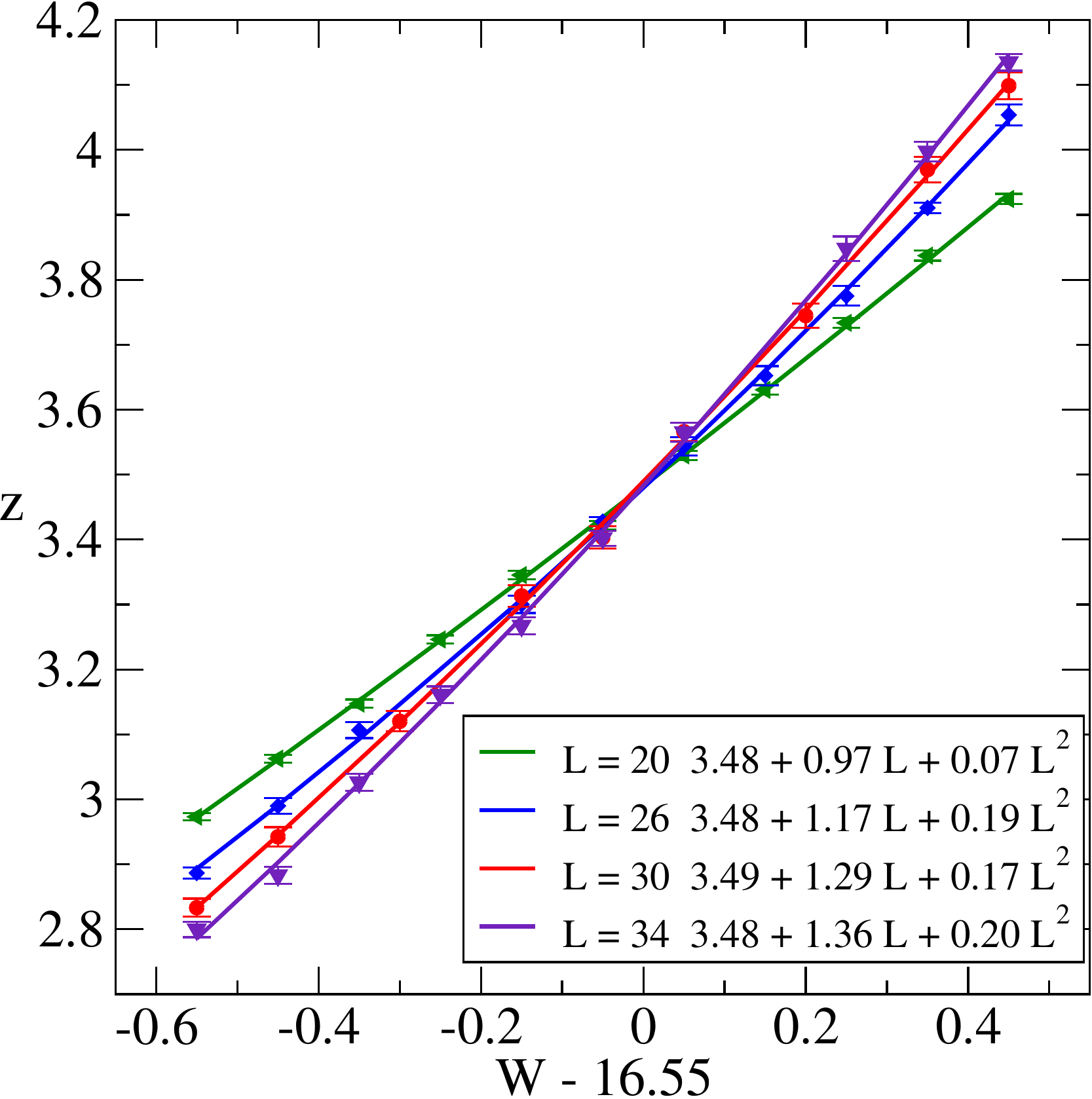}
\caption{Quadratic fit (\ref{ff}) of $z_1(W-W_c)$ for four values of the size $L$.
The cross section determines $z_{1c}\approx 3.48$.
} 
\label{v1_l30}
\end{figure}

Before testing the validity of  Eq. (\ref{s-1}) we have to notice the relation (\ref{eq-z1}) between the localization length expressed in Eq. (\ref{s-1}) and the parameter $z_1$ shown in Fig. \ref{w16-17}.
We fit our data for $z_1$ to  the function
\be\label{fit-s}
\zeta = \ds{\frac{1}{a_0+a_1L}}
\ee
shown by dotted lines in Fig. \ref{w16-17}. Comparing  with Eq. (\ref{s-1}) and using  $y^*= z_{1c}^{-1} = 3.48^{-1}$  (Fig. \ref{v1_l30}) we obtain
$L_0\approx 8.6$ from $W=16$ data, but significantly different value  $L_0\approx 17$ for $W=17$. 

\begin{figure}[t!]
\includegraphics[width=6cm]{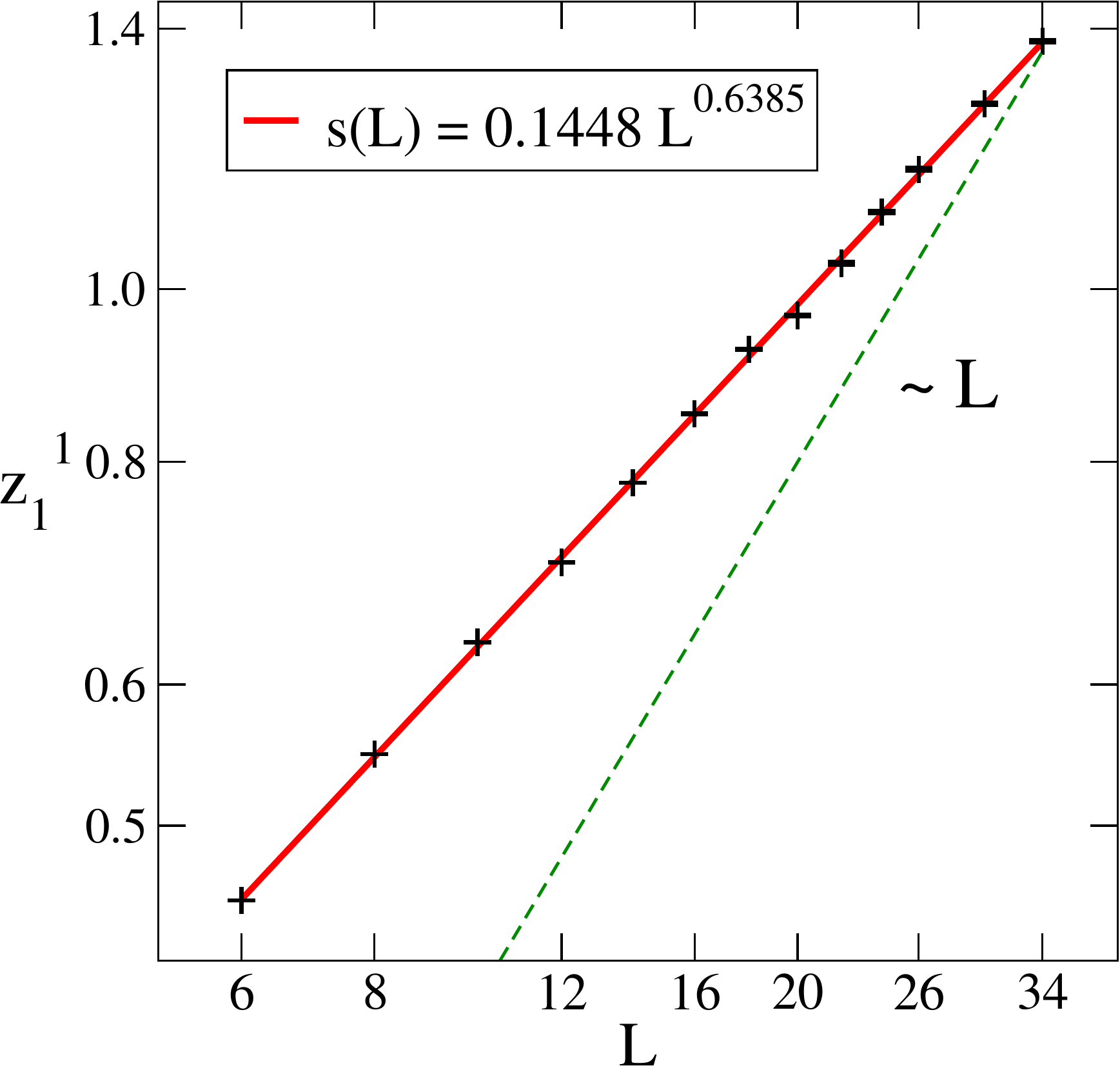}
%\vspace*{0.5cm}
\caption{The $L$-dependence of the slope $s(L) \sim L^{1/\nu}$.
The critical exponent $\nu = 1.566$. Dashed line shows  the linear $L$-dependence,
predicted by Eq. (\ref{s-1}). 
}
\label{z1}
\end{figure}

Although the power fit (\ref{s-1a}) is clearly better than the fit (\ref{fit-s}),
Fig. \ref{w16-17} shows that the estimation of true scaling behavior might be difficult since
various analytical functions seem to  fit numerical data with  sufficient accuracy.
 In the present case, the
problem lies in the non-zero critical value $z_{1c}$. 
To avoid the ambiguity in the choice of the fitting function,
we have to extract the critical  value from numerical  data\cite{jpa}.
When data for $z_1$ are plotted as a function of the disorder (Fig. \ref{v1_l30}), we can fit them by 
quadratic polynomial
\be\label{ff}
z_1(W,L) = z_{1c} + \tau s(L) + \tau^2 t(L)
\ee
and calculate the $L$-dependence of the slope $s(L)$. From Eq. (\ref{s-1}) we see that $s(L)$ should be 
a linear function of $L$, while Eq. (\ref{s-1a}) predicts power-law behavior $s(L)\sim L^{1/\nu}$.
Figure \ref{z1} shows $s(L)$ as a function of $L$.
The fit  confirms the power-law dependence $s(L)\sim L^{1/\nu}$ with  critical exponent
$\nu\approx  1.56$, as obtained by other methods \cite{sc}. 

\section{5D model}

For higher dimension, the following size dependence of the localization length 
at the critical point ($\tau=0$) was derived 
\be\label{f-5}
\frac{\xi_{\rm 1D}}{L} = \left(\ds{\frac{L}{a}}\right)^{(d-4)/3}.
\ee
In particular, for $d=5$ Eq. (\ref{f-5}) gives
\be
z_{1}(\tau=0) \sim L^{-1/3}
\ee
which means that  the critical value of $z_1$ is not size independent
but decreases to zero when  $L\to \infty$.
\begin{figure}[t!]
\includegraphics[width=6cm]{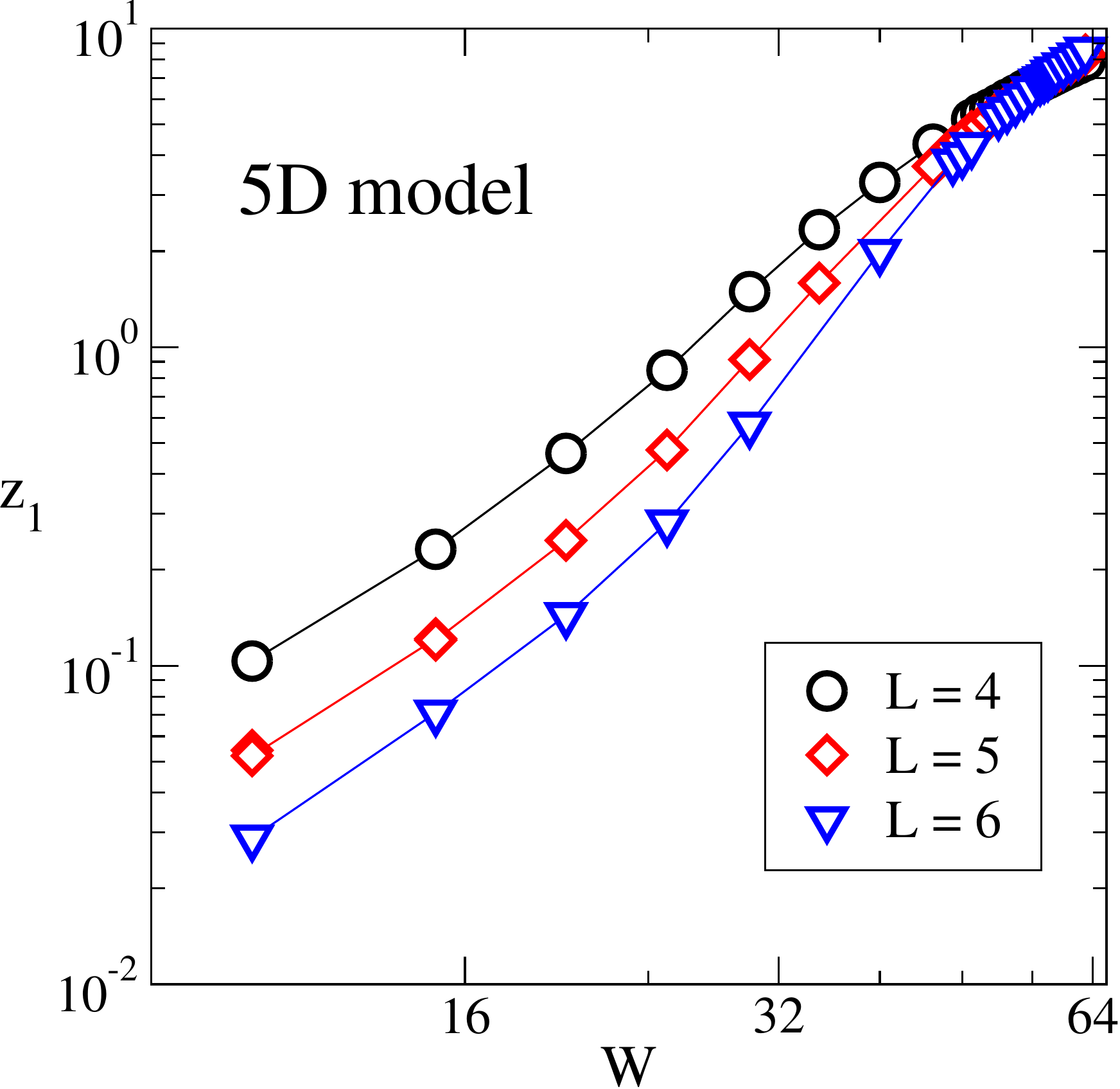}
%\vspace*{0.5cm}
\caption{The parameter $z_1(W)$ for various system size.
For $W< 57.5$,   $z_1$ decreases when $L$ increases.
There is no indication for the critical behavior described by 
Eq. (\ref{5d-crit}) 
}
\label{suslov-5d}
\end{figure}
Since the localization length is finite in for $\tau>0$,  the $\tau$-dependence of
$z_1(L,\tau)$  for fixed $L$ must exhibit an infinite discontinuity at $\tau=0$:
\be\label{5d-crit}
z_1(\tau) \sim\left\{
\begin{array}{ll}
L^{-1/3}  &   \tau=0\\
L         &    \tau>0
\end{array}
\right.
\ee
We test numerically the size and disorder dependence of $z_1$.
We show in Fig. \ref{suslov-5d} and \ref{5D} the disorder dependence of $z_1$
for fixed $L$. 
Our data in Fig. \ref{suslov-5d} do not indicate any discontinuity in the $L$ dependence. Contrary,
$z_1$ is smooth analytical function of both parameters, $W$ and $L$.

For smaller disorder, $z_1$ is always decreasing function of $L$. This is typical for
the  metallic regime. 
However,  $z_1$ does not depend on the size $L$ when
$W=57.5$. This is consistent with   the scaling equation (\ref{s-1a}).
Insulating regime, where $z_1$ increases with the size $L$ is observed only when $W>57.5$
(Fig. \ref{5D}).

Note that $z_1\approx 7$ for disorder $W\approx 57.5$. Therefore the localization length, 
\be
\xi_{\rm 1D} = \ds{\frac{2}{z_1}L}
\ee
is much smaller than the size of the system and  we do not expect that finite size effects
play significant role 
although the size $L$ is much smaller than in 3D system.

Scaling analysis, similar to that for the 3D model enables us to find the critical exponent, $\nu_{\rm 5D}\approx 0.96$.

%In Fig. \ref{suslov-5d} we present the disorder dependence of $z_1$ for $W\ge 8$. 
%and  $L=4, 5$ and 6. 
%Our data confirm  that $z_1$ is analytical function of both $W$ and $L$.
%No indication for the discontinuity (\ref{5d-crit}) is observable.
%Moreover, $z_1(L)$ decreases with increasing $L$ for all values of the disorder $W<57$
%indicating  that the system is in the metallic regime. 
%The localized regime is found for $W>57$ (Fig. \ref{suslov-5d}). Note that
%the size and disorder dependence of
% $z_1$ agrees with Eq. (\ref{s-1a}). 

\begin{figure}[t!]
\includegraphics[width=6cm]{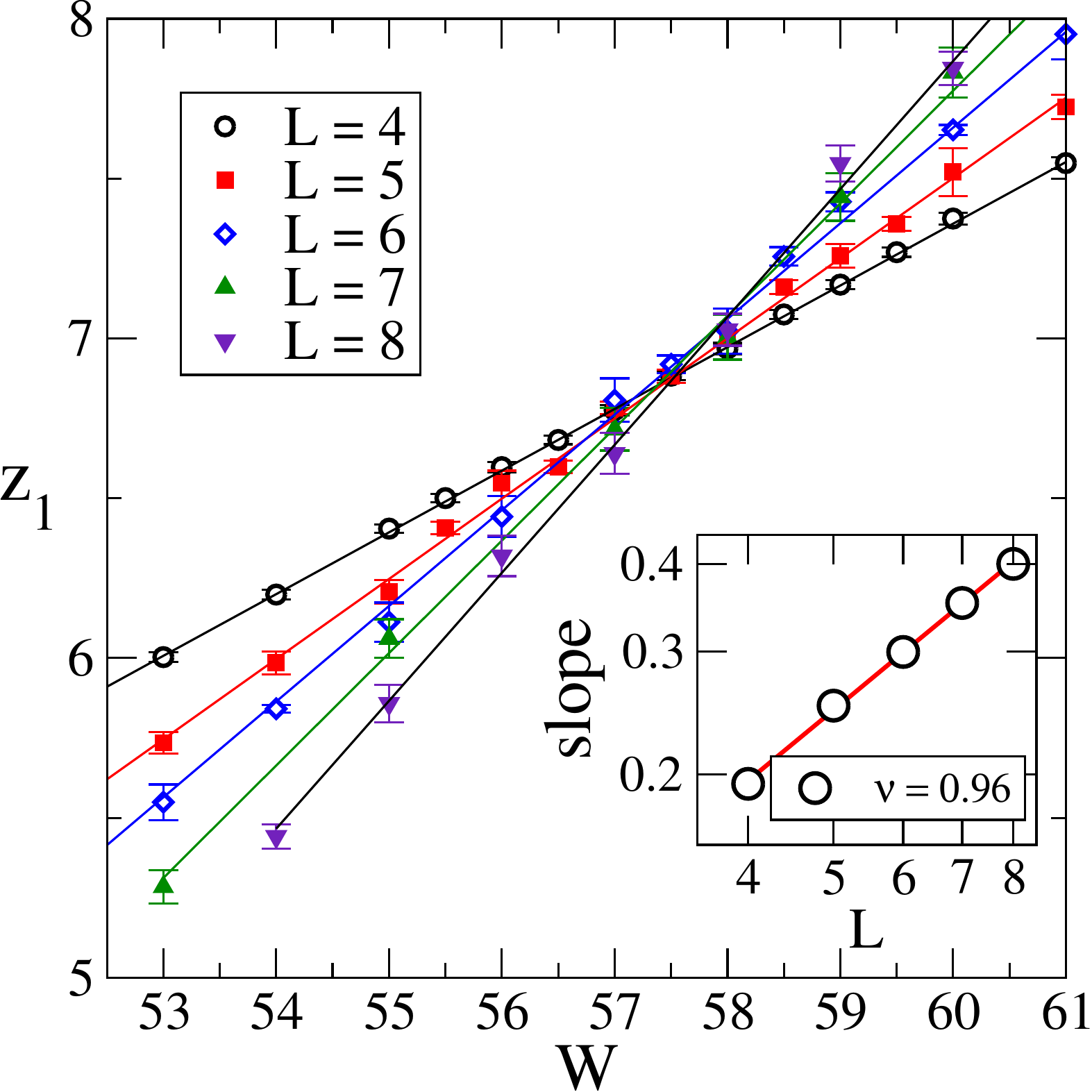}
%\vspace*{0.5cm}
\caption{The 5D Anderson model:
the parameter $z_1$ as a function of disorder $W$ for $L=4,5,6,7$ and $L=8$.
Data indicate that $z_1$ does not depend on the size $L$ when $W\approx 57.5$.
This value is considered as a critical disorder $W_c$ in the ``standard'' finite
size scaling theory. Solid lines are fits $z_1(L) = z_1+ s(L)(W-W_c)$. Inset
shows the $L$-dependence of the slope $s(L) \sim L^{1.0413}$.
The original figure was published in \cite{2} but new data for $L=8$ were added.
}
\label{5D}
\end{figure}

\section{Conclusion}

We showed that numerical data for the parameter $z_1$ do not agree with the
predictions of the theory \cite{1}. Both $z_1$ and the localization length are analytical continuous functions of  the disorder $W$  and the size of the system $L$.  

For the 3D system, we presented additional numerical data for larger system size $L\le 24$  up to $L=34$. 
These new data confirm
previous estimation of the critical exponent $\nu=1.56$.\cite{so,jpa} 
It is worth to mention that the same value of the critical exponent   was
obtained already 20 years ago with the use of numerical data for $L\le 12$ only \cite{mac}.
We also note that the same value of the critical exponent  
 was obtained from numerical analysis of other physical quantities: mean conductance, 
conductance distribution, inverse participation
ratio \cite{2} and also  for critical points  outside the band center \cite{2,4}. 
This value of critical exponent was  recently verified experimentally
\cite{3}  and calculated analytically \cite{gg}.

\end{document}